\documentclass[3p,times,procedia]{elsarticle}
\UseRawInputEncoding
\usepackage{ecrc}
\usepackage[bookmarks=false]{hyperref}
    \hypersetup{colorlinks,
      linkcolor=blue,
      citecolor=blue,
      urlcolor=blue}

\usepackage{floatrow}
\newfloatcommand{capbtabbox}{table}[][\FBwidth]

\usepackage{blindtext}

\volume{00}

\firstpage{1}

\journalname{Procedia Computer Science}

\runauth{Georgescu et al.}

\jid{procs}

\usepackage{amssymb}
\usepackage{amsmath}
\usepackage{subcaption,booktabs}

\usepackage[figuresright]{rotating}

\begin{document}
\begin{frontmatter}



\dochead{27th International Conference on Knowledge-Based and Intelligent Information \& Engineering Systems (KES 2023)}%

\title{Masked Autoencoders for Unsupervised  Anomaly Detection in Medical Images}


\author[a]{Mariana-Iuliana Georgescu}   

\address[a]{Faculty of Mathematics and Computer Science, University of Bucharest, 14 Academiei, Bucharest, Romania} 

\begin{abstract} 
Pathological anomalies exhibit diverse appearances in medical imaging, making it difficult to collect and annotate a representative amount of data required to train deep learning models in a supervised setting. Therefore, in this work, we tackle anomaly detection in medical images training our framework using only healthy samples. We propose to use the Masked Autoencoder model to learn the structure of the normal samples, then train an anomaly classifier on top of the difference between the original image and the reconstruction provided by the masked autoencoder. We train the anomaly classifier in a supervised manner using as negative samples the reconstruction of the healthy scans, while as positive samples, we use pseudo-abnormal scans obtained via our novel pseudo-abnormal module. The pseudo-abnormal module alters the reconstruction of the normal samples by changing the intensity of several regions. We conduct experiments on two medical image data sets, namely BRATS2020 and LUNA16 and compare our method with four state-of-the-art anomaly detection frameworks, namely AST, RD4AD, AnoVAEGAN and f-AnoGAN. 

\end{abstract}

\begin{keyword}
anomaly detection \sep medical imaging \sep self-supervised learning \sep deep learning \sep masked autoencoders




\end{keyword}
\end{frontmatter}

\email{georgescu{\_}lily@yahoo.com}



\vspace{-1.2\baselineskip}
\section{Introduction}
\label{sec:intro} 
\vspace{-0.5\baselineskip}
Masked data modeling which employs the Transformer architecture has been vastly adopted by the deep learning community ranging from text~\cite{Devlin-ACL-2019} to images~\cite{He-CVPR-2022}, videos~\cite{Tong-NIPS-2022} and sound~\cite{Huang-NIPS-2022}. The masked autoencoder models were mainly used for pre-training, due to their ability to leverage the structure already existing in data. 
Consequently, in this work, we employ them for anomaly detection in medical images to learn the underlying structure of normal data. 

Meanwhile, the medical imaging domain has witnessed significant advancement with applications ranging from medical image super-resolution to organ segmentation, disease progress prediction, tumor detection, etc. Researchers have invested a lot of effort in applying deep learning methods in medical imaging, obtaining outstanding results using supervised learning for tumor detection~\cite{Zheshu-ACCESS-2020,Mohsen-FCIJ-2018,Song-NC-2020}, organ segmentation, cancer prediction, etc. Even though supervised methods obtained noteworthy results, the acquisition of supervised annotations for medical applications remains a challenging task, requiring specialized expertise. However, there is a lot of unlabeled data which can be used to train deep learning models. In order to leverage the available unlabeled data, researchers~\cite{Baur-MICCAIW-2019,Baur-MICCAI-2020,Bengs-MICAD-2022,Chen-MIDL-2018,Chen-MIA-2020,Kascenas-MIDL-2022,Pinaya-MIA-2022,Wolleb-MICCAI-2022,Zimmerer-MIDL-2019} have focused on unsupervised anomaly detection in medical images, where a model is trained using only normal data (these techniques are sometimes considered semi-supervised). Researchers trained Autoencoders~\cite{Baur-MICCAI-2020,Chen-MIDL-2018, Kascenas-MIDL-2022,Shvetsova-ACCESS-2021}, Variational Autoencoders~\cite{Baur-MICCAIW-2019,Pinaya-MIA-2022, Zimmerer-MIDL-2019}, Generative Adversarial Networks~\cite{Han-BMCB-2021,Schlegl-IPMI-2017, Schlegl-MIA-2019} and Denoising Diffusion Models~\cite{Pinaya-MICCAI-2022, Wolleb-MICCAI-2022} to capture the structure of normal data. Their approaches are based on the concept that the models should reconstruct a normal sample with high fidelity, but they should face difficulties in accurately reconstructing an abnormal instance. Therefore, the anomaly score is computed using a function that measures the difference between the original image and the reconstructed image. Using this approach of computing the anomaly score, the anomaly decision is only based on the reconstruction fidelity of the underlying model. To overcome the limitation of relying entirely on the reconstruction fidelity of the model, we propose an anomaly classifier to predict the probability of a sample to be abnormal. Our anomaly classifier is trained using supervised learning.

In this work, we learn informative and discriminative features from medical images under weak annotations by using only normal data (healthy scans) at training time. We employ the masked autoencoder model to learn the underlying patterns of data in a self-supervised manner. The masked autoencoder model reconstructs a given image using only a small portion (usually $25\%$) of its pixels, thus during training it must capture the structure of data in order to accurately reconstruct the masked input. Training the masked autoencoder model only using healthy scans, forces the model to reconstruct with high fidelity only normal patterns, therefore it should face difficulties in accurately reconstructing an abnormal sample. We then propose to train an anomaly classifier in a supervised fashion. To obtain abnormal training data, we apply a novel pseudo-abnormal module to alter the reconstruction of the normal samples obtaining pseudo-abnormal examples.
We test our method on two medical image data sets, namely BRATS2020~\cite{Menze-TMI-2015} and LUNA16~\cite{Setio-MIA-2017}, outperforming four state-of-the-art anomaly detection frameworks~\cite{Baur-MICCAIW-2019,Deng-CVPR-2022,Rudolph-WACV-2023,Schlegl-MIA-2019}.
 
To the best of our knowledge, our contributions are twofold:
\vspace{-0.8\baselineskip}
\begin{itemize} 
    \item We apply the MAE framework for anomaly detection in medical imaging.  
    \item We proposed an anomaly classifier to further boost the performance of frameworks which rely on the reconstruction fidelity of the underlying model.
\end{itemize} 

\vspace{-2.0\baselineskip}
\section{Related Work} 
\vspace{-0.8\baselineskip}
Due to the success of unsupervised and self-supervised algorithms in computer vision applications, many researchers have started to apply unsupervised algorithms in medical imaging. Baur et al.~\cite{Baur-MIA-2021} presented in their survey study that the works that tackle the medical image anomaly detection can be divided into three categories based on the architecture. The architectures used are: Autoencoder (AE)~\cite{Baur-MICCAI-2020,Chen-MIDL-2018, Kascenas-MIDL-2022,Shvetsova-ACCESS-2021}, Variational Autoencoder (VAE) \cite{Baur-MICCAIW-2019, Zimmerer-MIDL-2019} and Generative Adversarial Networks (GAN) \cite{Han-BMCB-2021,Schlegl-IPMI-2017,Schlegl-MIA-2019}. With respect to the addressed task, related works can be divided into methods that perform anomaly detection (i.e. the model only predicts if an image is abnormal)~\cite{Bengs-MICAD-2022,Zimmerer-MIDL-2019} or methods that perform anomaly segmentation (i.e. the model also localizes the anomalous regions)~\cite{Baur-MICCAIW-2019,Baur-MICCAI-2020,Chen-MIDL-2018,Chen-MIA-2020,Kascenas-MIDL-2022,Pinaya-MIA-2022,Wolleb-MICCAI-2022}. 

Similar to~\cite{Bengs-MICAD-2022,Zimmerer-MIDL-2019}, we propose an unsupervised algorithm that performs anomaly detection in medical images.

Chen et al.~\cite{Chen-MIDL-2018} proposed an algorithm to detect the anomalies in MRI brain scans by adding an adversarial constraint on the latent space of the AE model. The constraint enforces that an abnormal image be mapped in the same point as a normal image, such that the difference between the reconstruction of an abnormal image and the original image will highlight the lesion. To apply the constraint, the algorithm requires adversarial (abnormal) data. Chen et al.~\cite{Chen-MIDL-2018} used as adversarial samples, the reconstructions of the healthy samples. Different from the other works that employ AE to segment anomalies in medical images, Baur et al.~\cite{Baur-MICCAI-2020} learned to compress and reconstruct different frequency bands of healthy brain MRI scans using the Laplacian pyramid.
Kascenas et al.~\cite{Kascenas-MIDL-2022} showed that a simple yet effective, denoising autoencoder architecture obtains better results than more sophisticated methods~\cite{Zimmerer-MIDL-2019} on the brain lesion segmentation task.

To improve the anomaly segmentation performance, Baur et al.~\cite{Baur-MICCAIW-2019} proposed AnoVAEGAN.  AnoVAEGAN  trained the decoder model with the help of an adversarial network, which discriminates between real images and images reconstructed by the VAE model. 

Instead of treating the anomaly detection task as an image reconstruction task, Chen et al.~\cite{Chen-MIA-2020} approached the unsupervised lesion detection as an image restoration problem proposing a probabilistic model to detect brain lesions.  

Different from the aforementioned works~\cite{Baur-MICCAIW-2019,Baur-MICCAI-2020,Chen-MIDL-2018,Han-BMCB-2021, Kascenas-MIDL-2022, Pinaya-MIA-2022, Schlegl-IPMI-2017,Shvetsova-ACCESS-2021,Zimmerer-MIDL-2019} which only use the MRI scans as input to detect the anomalies, Bengs et al.~\cite{Bengs-MICAD-2022} proposed a model for unsupervised 3D brain MRI anomaly detection considering the age as additional information.  Bengs et al.~\cite{Bengs-MICAD-2022}  considered age prediction as an additional task and attached a regression layer, just before the latent space of the VAE model, which predicted the age of the patient. Bengs et al.~\cite{Bengs-MICAD-2022} observed that the prediction error of the age was increased by a factor of two on abnormal data.

More recently, generative adversarial networks have been surpassed by the denoising diffusion probabilistic models in the computer vision community. \cite{Pinaya-MICCAI-2022, Wolleb-MICCAI-2022} proposed to use a denoising diffusion probabilistic model to segment anomalies in brain imaging. Wolleb et al.~\cite{Wolleb-MICCAI-2022} presented a novel pixel-wise anomaly detection approach based on Denoising Diffusion Implicit Models. 

Pinaya et al.~\cite{Pinaya-MIA-2022} proposed to employ a Transformer model to learn the probability density function only from healthy brain scans.  Pinaya et al.~\cite{Pinaya-MIA-2022} integrated the Transformer into the vector quantized variational autoencoder model showing the benefits of using a self-attention model on the brain anomaly segmentation task. Instead of training the model to reconstruct the input, Pinaya et al.~\cite{Pinaya-MIA-2022} optimized the model towards outputting the likelihood of the input. 

Most of the aforementioned works~\cite{Baur-MICCAIW-2019,Baur-MICCAI-2020,Chen-MIDL-2018, Han-BMCB-2021, Kascenas-MIDL-2022, Schlegl-IPMI-2017, Shvetsova-ACCESS-2021,Zimmerer-MIDL-2019} rely on the reconstruction fidelity to compute the anomaly score. Different from such works, we employ a Transformer model, which is trained in a supervised manner, to discriminate between normal and abnormal reconstructions, alleviating the limitation of relying only on the reconstruction fidelity of the underlying model. Similar to~\cite{Chen-MIDL-2018}, we use adversarial samples, but we build our own pseudo-abnormal sample pool instead of using the reconstructions of the samples as abnormal examples. 
Similar to~\cite{Pinaya-MIA-2022}, our underlying model is a Transformer. Different from~\cite{Pinaya-MIA-2022}, we employ the Masked Autoencoder (MAE) framework to learn discriminative features from healthy scans. To the best of our knowledge, we are the first to propose the use of masked autoencoders to detect anomalies in medical images.

\vspace{-0.5\baselineskip} 
\section{Method}
\vspace{-0.5\baselineskip} 
We begin with an overview of the proposed method (Section~\ref{overview}), then we describe each component of our approach, namely the Masked Autoencoder framework (Section~\ref{mae}), the Pseudo-abnormal Module (Section~\ref{pseudo-abnormal module}) and the Anomaly Classifier (Section~\ref{anomaly_classifier}).

\subsection{Overview}
\label{overview}
\begin{figure*}[t]
\begin{center}
\includegraphics[width=0.99\linewidth]{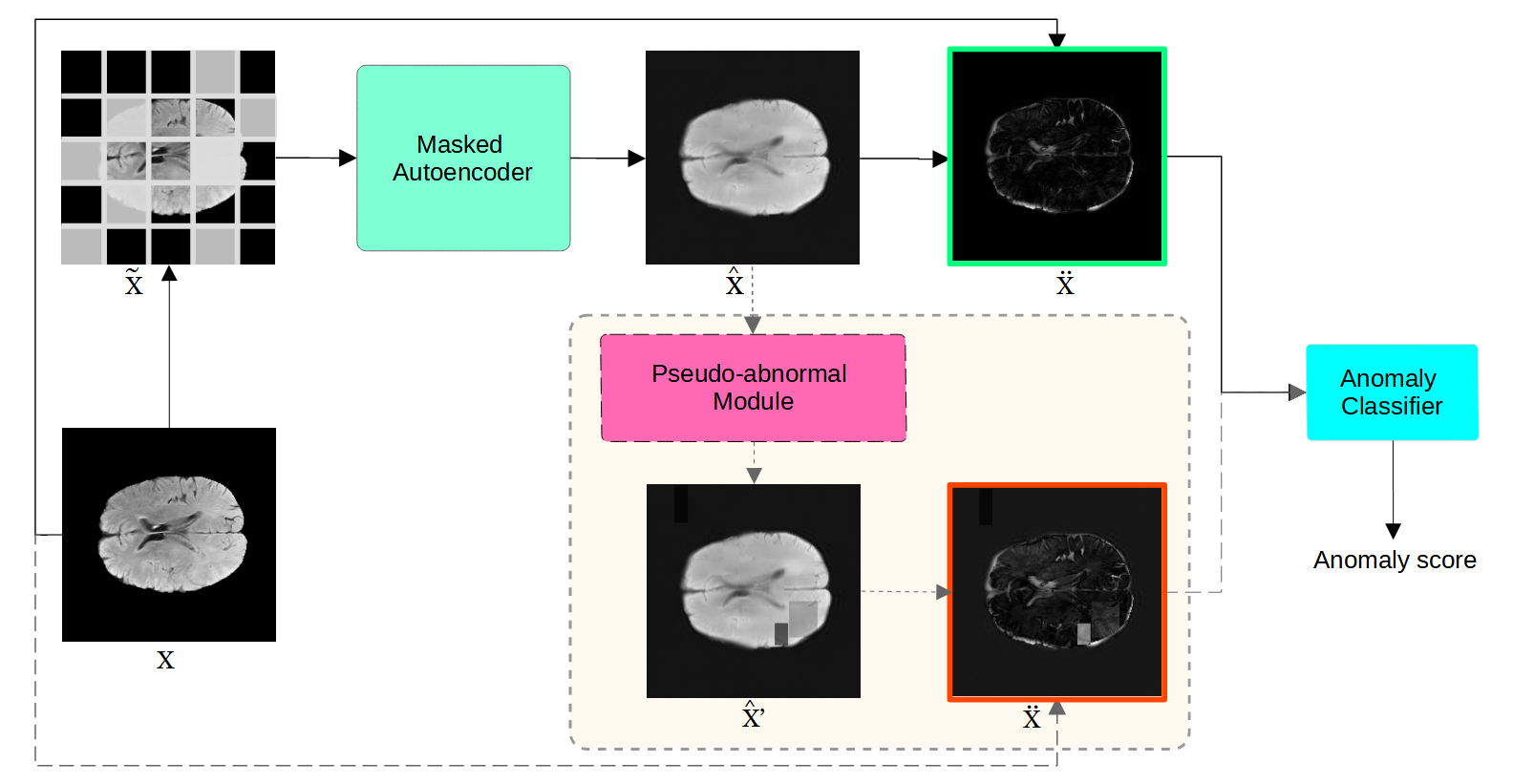}
\end{center}
\caption{Overview of our framework for unsupervised anomaly detection in medical images. We first train the MAE model using only healthy samples, designated as $X$. We use a pseudo-abnormal module to alter $\hat{X}$, the reconstruction provided by the MAE module, in order to obtain $\hat{X'}$, the pseudo-abnormal samples utilized to train the anomaly classifier in a supervised manner. The anomaly classifier outputs the probability of a given sample to be abnormal based on $\Ddot{X}$, the difference between the reconstructed and the original sample. The components depicted in dashed lines are not used during inference.}\label{fig_pipeline}
\end{figure*}

\vspace{-0.5\baselineskip}
We illustrate our approach of detecting anomalies in medical images in Figure~\ref{fig_pipeline}. We start by applying the Masked Autoencoder framework (MAE) to obtain the reconstruction of the medical image input. Afterwards, the absolute difference between the reconstructed input and the actual input is used to train the anomaly classifier. The binary anomaly classifier is trained in a supervised manner requiring both positive and negative examples. As negative examples, we use the reconstructed images, while for the positive class, we generate pseudo-abnormal samples.
We use a pseudo-abnormal module which is applied on the reconstructed input to create a set of pseudo-abnormal samples. The anomaly score for a sample is given by the probability of being labeled as pseudo-abnormal by the anomaly classifier.  

\subsection{Masked Autoencoders}
\vspace{-0.5\baselineskip} 
\label{mae}
Transformer models have recently received a lot of attention from the deep learning community and have started being the standard architecture used in computer vision and natural language processing models. According to Shamshad et al.~\cite{Shamshad-arxiv-2022}, Vision Transformer models (ViT)~\cite{Dosovitskiy-ICLR-2021} are heavily used in medical imaging processing too. Transformers use as input a 1D sequence of token embeddings. In order to apply a Transformer model on an image, the image is divided into patches, then each patch is projected into a 1D embedding space forming the tokens.

Transformer models need huge amounts of training data to avoid overfitting. To overcome this challenge, He et al.~\cite{He-CVPR-2022} proposed the masked autoencoder framework, a self-supervised method to pre-train the ViT model on small data sets. 
He et al.~\cite{He-CVPR-2022} designed an asymmetric encoder-decoder model that operates on patches. The key concept of the MAE framework is to  randomly remove tokens from the input, then process the remaining tokens with the encoder, and finally reconstruct the unseen tokens with a lightweight decoder from the latent representation and the learned mask tokens. 

In this work, we do not employ the MAE framework as a self-supervised method, but we employ it for its capability to learn the structure of the data set. We train the MAE framework using only normal data in order to capture the patterns of normal samples.

The MAE framework~\cite{He-CVPR-2022} works as follows. Let $X$ be the input image and $\Tilde{X}$ be the version of the input $X$, which is divided into patches. $\Tilde{X}$ is transformed into embedding tokens through a linear projection obtaining $\Tilde{P}=[p_1, p_2, .., p_n]$ where $\Tilde{P} \in  \mathcal{R} ^ {n\times d}$, $n$ is the number of patches and $d$ is the embedding dimension. Additionally, a positional embedding is added to the embeddings input $\Tilde{P}$. The token sequence becomes $P = \Tilde{P} + pos$, where $pos$ is the positional embedding. 

\begin{figure*}[!t]
\begin{center}
\includegraphics[width=0.99\linewidth]{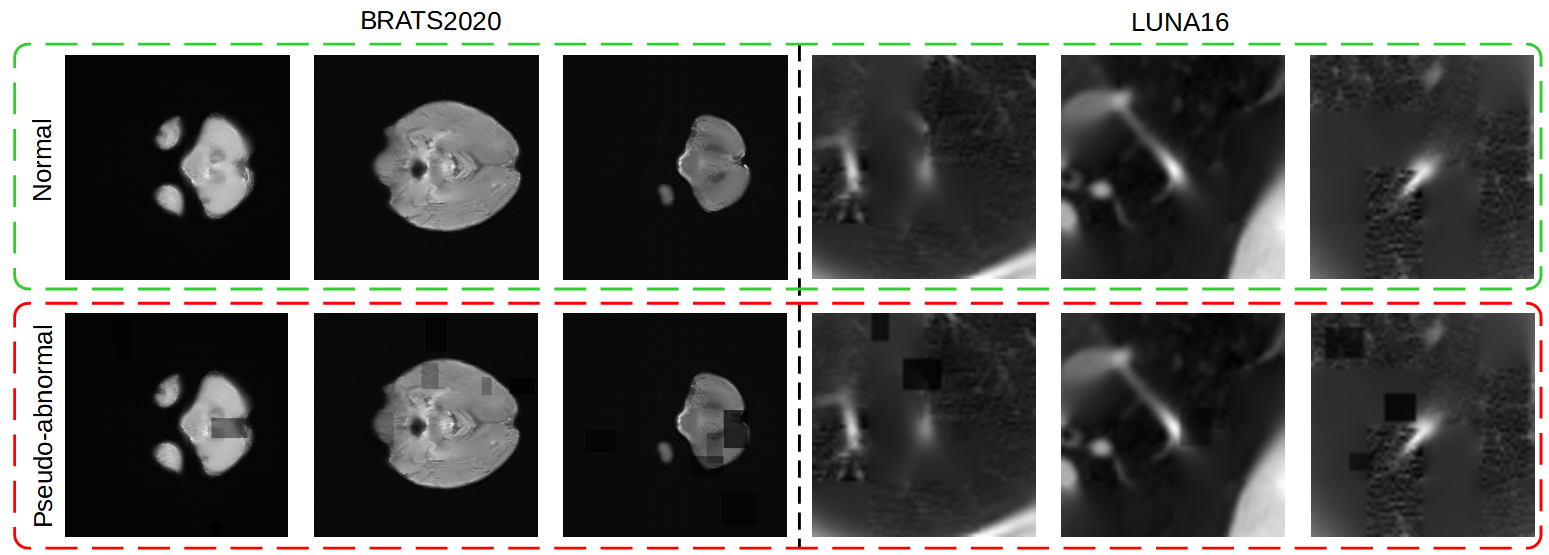}
\end{center}
\caption{Anomaly classifier training samples extracted from the BRATS2020~\cite{Menze-TMI-2015} and LUNA16~\cite{Setio-MIA-2017} data sets. The normal samples are the reconstructed images obtained using MAE, while the pseudo-abnormal samples are the reconstructed images after applying the pseudo-abnormal module. The samples are employed to train the anomaly classifier.}\label{fig_samples}
\end{figure*}

A random subset of tokens from $P$ are removed (masked) and the remaining (visible) tokens are processed by the encoder obtaining the encoded representation. More formally, this operation can be expressed as $V = mask(P, \alpha)$ and $E = encoder(V)$, where $mask$ is the masking operation where the tokens are dropped and $\alpha$ is the masking ratio.

The last component of the MAE framework is the lightweight decoder. The decoder is applied on top of the encoded representation $E$ and the learned mask tokens. The learned mask tokens $m \in \mathcal{R} ^ {d}$ are placed in the location of the unseen tokens forming the encoded token sequence $\Bar{Z} \in  \mathcal{R} ^ {n \times d}$. A new positional embedding is added to the encoded token sequence $\Bar{Z}$, resulting $Z = \Bar{Z} + pos_d$, where $pos_d$ is the positional embedding. Eventually, the reconstruction is obtained as $\hat{P} = decoder(Z)$. By re-arranging the tokens $\hat{P}$, we obtain the reconstruction $\hat{X}$ of the input image $X$.

The MAE framework is optimized by employing the mean squared error loss between the reconstructed tokens $\hat{P}$ and the input tokens $\Tilde{X}$ in the pixel space. The employed loss function is:
\vspace{-1\baselineskip}
\begin{equation}
\mathcal L(\Tilde{X}, \hat{P}) = \frac{1}{\alpha \cdot n} \sum_{i \in \mathcal{M}} || \Tilde{x}_i - \hat{p}_i ||^{2},
\label{eq:reconstruction_mae}
\vspace{-1.5\baselineskip}
\end{equation}  
where $\mathcal{M}$ denotes the set of masked token indices, as we only compute the reconstruction loss on the masked tokens. In Section~\ref{sec_exp_ablation}, we present an ablation experiment showing that the anomaly detection performance is higher when we apply the loss only on the masked tokens.

\vspace{-0.3\baselineskip}
\subsection{Pseudo-abnormal Module}   
\vspace{-0.3\baselineskip} 
\label{pseudo-abnormal module}
 
In order to train the anomaly classifier in a supervised setting, both positive and negative samples are required. As negative (\textit{normal}) samples, we use the reconstructed images. To overcome the limitation of not having real positive (\textit{abnormal}) samples, we create \textit{pseudo-abnormal} samples which we use as positive samples during training. In order to create positive samples, we intentionally alter the reconstructed images. The concept behind this design choice is that if an abnormal example is presented during inference, the reconstructed image should exhibit artifacts, therefore, we intentionally add artifacts to the reconstructed image. For each negative reconstructed image $\hat{X}$, we produce a positive sample $\hat{X'}$. To create $\hat{X'}$, we first select $k$ random bounding boxes of different dimensions inside $\hat{X}$. For each selected bounding box, we change the intensity of the pixels inside the bounding box by multiplying them by $\beta$. The multiplier factor $\beta$ is randomly selected from the uniform distribution $\textit{\textbf{U}}(0, 1)$.

Our design choice of altering the reconstruction using randomly generated boxes is based on the concept that the tokens (image patches) which are included in an abnormal region should not be accurately reconstructed by the MAE framework. Therefore, we simulate this scenario by altering random patches from the reconstructed image.

In Figure~\ref{fig_samples}, we illustrate a few examples of positive and negative samples used as training data for the anomaly classifier.

\vspace{-0.75\baselineskip}
\subsection{Anomaly Classifier} 
\label{anomaly_classifier}
\vspace{-0.5\baselineskip} 
The last component of our method is the Anomaly Classifier. We employ the ViT model as the underlying architecture. We train the classifier to distinguish between positive (\textit{pseudo-abnormal}) and negative (\textit{normal}) samples. The input to the classifier, denoted as $\Ddot{X}$, is the absolute difference between the reconstructed sample $\hat{X}$ (for negative) or $\hat{X'}$  (for positive) and the original image $X$. 

The classifier is optimized using the binary cross-entropy function:
\begin{equation}\label{eq_BCE_loss}
\mathcal{L}_{\mbox{\scriptsize{cross-entropy}}}(y, \hat{y}) = - y \cdot \log(\hat{y}) + (1 - y) \cdot \log(1 - \hat{y}),
\end{equation}
where $y$ is the ground-truth label ($0$ for negative samples and $1$ for positive samples) and $\hat{y}$ represents the prediction of sample $\Ddot{X}$. 

At inference time, when we have both normal and abnormal samples, the anomaly score is interpreted as the probability of a sample to be positive (pseudo-abnormal).

\section{Experiments}  
We start by describing our experimental setup (Sections~\ref{datasets},~\ref{eval_metrics} and~\ref{implementation_details}), then we compare our approach with state-of-the-art methods in Section~\ref{results} and present our ablation study in Section~\ref{sec_exp_ablation}.

\vspace{2\baselineskip}
\begin{minipage}{.41\textwidth}
 \captionof{table}{Anomaly detection results in terms of AUROC \\ on  the BRATS2020~\cite{Menze-TMI-2015} and LUNA16~\cite{Setio-MIA-2017} data sets. \\ The top results are highlighted in bold.}
    \label{table_results} 
    \scalebox{0.95}{
    \begin{tabular}{lll}
    \multicolumn{1}{c}{\bf Method}  &\multicolumn{2}{c}{\bf AUROC} \\
    \cline{2-3}
                         & BRATS2020 & LUNA16
    \\ \hline   
    AnoVAEGAN~\cite{Baur-MICCAIW-2019}     &  $0.872$ &  $0.583$\\\
    f-AnoGAN~\cite{Schlegl-MIA-2019}     &  $0.863$ &  $0.535$\\
    RD4AD~\cite{Deng-CVPR-2022}      &  $0.886$ &  $0.521$\\ 
    AST~\cite{Rudolph-WACV-2023}     &  $0.895$ &  $0.619$\\
    Proposed method                  &  $\mathbf{0.899}$ &  $\mathbf{0.634}$\\
    \end{tabular}
    }
\end{minipage}
\hspace{2\baselineskip}
\begin{minipage}{.5\textwidth} 
 \vspace{0.2\baselineskip}
 
  \includegraphics[width=1.0\linewidth]{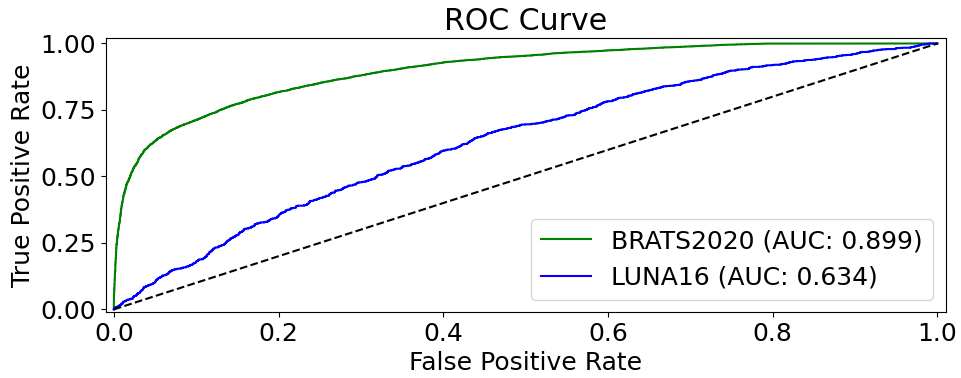}
    \vspace{-1.5\baselineskip}
   \captionof{figure}{The ROC curves obtained by our proposed method on the BRATS2020~\cite{Menze-TMI-2015} and LUNA16~\cite{Setio-MIA-2017} data sets. The dashed line is the ROC curve obtained by a random classifier.}\label{roc_curves}
\end{minipage}  

\subsection{Data Sets} \label{datasets}
\noindent 
\textbf{BRATS2020.} Multimodal Brain Tumor Segmentation Challenge 2020~\cite{Bakas-SD-2017,Bakas-arxiv-2018,Menze-TMI-2015} is an MRI data set containing $369$ scans. The tumors presented in the data set are glioblastoma and lower grade glioma. One of the main tasks performed on this data set is tumor segmentation. In this work, we tackle an easier problem, namely tumor detection. We treat the tumors as anomalies, but we train our model using only healthy scans. We annotate a slice as normal if there is no region containing tumor, otherwise we consider it as abnormal. We split the $369$ scans into $60\%$ scans for training and $40\%$ for test. We keep $20\%$ of the training scans for validation. We apply our algorithm at slice level, predicting if each slice is anomalous. After splitting the data set, we obtain $15,\!557$, $4,\!013$, $13,\!203$ normal slices for training, validation, and test, respectively, as well as $2,\!962$, $9,\!737$ abnormal slices for validation and test. The resolution of each scan is $240 \times 240$ pixels. In our experiments, we use only the native modality to determine if a slice is abnormal.

\noindent
\textbf{LUNA16.} LUng Nodule Analysis-2016~\cite{Setio-MIA-2017} contains $888$ lung CT scans with some of the scans containing pulmonary nodules. The data set provides a list of $551,\!065$ 2D regions which are candidates for being pulmonary nodules. One of the main tasks is the false positive reduction track, where the nodule candidates should be classified accordingly. Out of $551,\!065$ candidate regions, only $1,\!351$ are labeled as pulmonary nodules. In this work, we consider the pulmonary nodules as anomalies. We keep all abnormal regions for validation and testing and use only non-nodule regions for training. We randomly select $50,\!000$ out of $551,\!065$  non-nodule regions for training. We randomly select $3,\!000$ and $10,\!000$ normal regions for validation and test, while splitting the abnormal regions into $351$ and $1,\!000$ regions for validation and test, respectively. The resolution of each region is $64 \times 64$ pixels.
 
To facilitate further comparisons, we release the split of each data set used in our experiments at https://github.com/lilygeorgescu/MAE-medical-anomaly-detection.

\vspace{-0.5\baselineskip}
\subsection{Evaluation Metrics} \label{eval_metrics}
\vspace{-0.5\baselineskip}

As the evaluation metric, we employ the Area Under the Receiver Operating Characteristic curve (AUROC). The AUROC is calculated as the area under the receiver operating characteristic (ROC) curve. A ROC curve shows the trade-off between the true positive rate and false positive rate across different decision thresholds. The true positive rate is the proportion of samples correctly classified as abnormal, while the false positive rate is the proportion of normal samples misclassified as abnormal. The AUROC ranges from 0 to 1, with higher values indicating better performance.

\vspace{-0.3\baselineskip}
\subsection{Implementation Details} \label{implementation_details}
\vspace{-0.5\baselineskip}
In order to train the MAE framework, we use the official PyTorch implementation\footnote{\url{https://github.com/facebookresearch/mae}} provided by the authors~\cite{He-CVPR-2022}. We use ViT-Base~\cite{Dosovitskiy-ICLR-2021} as encoder for the MAE framework. The decoder architecture follows the setting proposed by He et al.~\cite{He-CVPR-2022} which has the embeddings dimension equal to $512$, the number of Transformer blocks set to $8$, and $16$ attention heads. We train the autoencoder only using normal samples for $1600$ epochs, setting the mask ratio to $0.75$. The input size is set to $224 \times 224 \times 1$ for the BRATS2020 data set, while for the LUNA16 benchmark, we set the input size to $64 \times 64 \times 1$. In order to obtain the reconstruction of a sample, we replaced the unmasked tokens with the original tokens. To obtain the final reconstruction of a sample, we pass it through the MAE framework $4$ times and average the resulting outputs. 

The number of bounding boxes ($k$) for the pseudo-abnormal module is uniformly sampled between $1$ and $10$. The width and height  of a bounding box are also randomly selected between $10$ and $40$ pixels for BRATS2020 and between $5$ and $12$  pixels for LUNA16, respectively.

The underlying architecture of the anomaly classifier is also ViT-Base~\cite{Dosovitskiy-ICLR-2021}. We train the anomaly classifier to discriminate between normal and pseudo-abnormal samples for $100$ epochs. The model is optimized using AdamW~\cite{Loshchilov-ICLR-2019} with the learning rate set to $0.001$ and the weight decay set to $0.05$.

The experiments are performed on a single GeForce GTX 3090 GPU with $24$ GB of VRAM. Our code is available online at https://github.com/lilygeorgescu/MAE-medical-anomaly-detection.

\vspace{-0.5\baselineskip}
\subsection{Results}  \label{results}
\vspace{-0.5\baselineskip}


\begin{table*}[t]
	\caption{
		Ablation results performed on the BRATS2020~\cite{Menze-TMI-2015} data set.
	}
	\begin{subtable}{0.38\linewidth}
		\caption{
			After removing the anomaly classifier component, we compute the AUROC score of the MAE component using as anomaly score the mean squared error, mean absolute error or the SSIM value between the reconstructed image and the original image.
		}
		\centering
		\scalebox{0.95}{\label{ablation_method}
            \begin{tabular}{ll} 
            \multicolumn{1}{c}{\bf Anomaly Measure}  &\multicolumn{1}{c}{\bf AUROC} \\
             \hline 
            Mean squared error          &  $0.614$\\
            Mean absolute error         &  $0.834$\\
            SSIM                        &    $0.877$\\
            \end{tabular}
		} 
	\end{subtable} \hfill
	\begin{subtable}{0.32\linewidth}
		\caption{
			 We vary the input of the anomaly classifier switching among: absolute difference ($L_1$), Euclidean distance ($L_2$) between the reconstructed and the original image, and the reconstructed image. \\    
		} 
		\centering
		\scalebox{0.95}{\label{ablation_input}
                \begin{tabular}{ll} 
                \multicolumn{1}{c}{\bf Input}  &\multicolumn{1}{c}{\bf AUROC}
                \\ \hline 
                $L_1$ difference         &  $0.899$\\
                $L_2$ difference         &  $0.875$\\
                Reconstructed image      &  $0.821$\\
                \end{tabular}
		}
		 
	\end{subtable}   \hfill
	\begin{subtable}{0.25\linewidth}
		\caption{
	We vary  $k_1$ \text{and} $k_2$, the minimum and the maximum numbers from which $k$ (the number of abnormal regions in an example) is sampled. \\  
		} 
		\centering
		\scalebox{0.95}{\label{ablation_regions}
                \begin{tabular}{ll}
                \multicolumn{1}{c}{\bf $[\ k_1, k_2 ]\ $} & \multicolumn{1}{c}{\bf  AUROC}
                \\ \hline  
                (1, 10)  &  $0.899$ \\
                (5, 10)  &  $0.889$ \\
                (1, 5)   &  $0.901$ \\ 
                \end{tabular}
		}
		\label{tab:ablation_anomaly_classifier}
	\end{subtable} 
\end{table*}

We present the results of our proposed method along with other state-of-the-art methods~\cite{Baur-MICCAIW-2019,Deng-CVPR-2022,Rudolph-WACV-2023,Schlegl-MIA-2019} for image anomaly detection on the BRATS2020~\cite{Menze-TMI-2015} and LUNA16~\cite{Setio-MIA-2017} data sets in Table~\ref{table_results}. We also illustrate the ROC curves obtained by our proposed method in Figure~\ref{roc_curves}. We emphasize that the analysis is performed at the slice level, therefore the results reported in Table~\ref{table_results} are the slice-wise AUC scores. In order to compute the performance of AnoVAEGAN, f-AnoGAN, RD4AD and AST on the BRATS2020 and LUNA16 data sets, we use the official code released by the authors\footnote{\url{https://github.com/hq-deng/RD4AD}}$^,$\footnote{\url{https://github.com/marco-rudolph/ast}}$^,$\footnote{\url{https://github.com/StefanDenn3r/Unsupervised_Anomaly_Detection_Brain_MRI}}. On the BRATS2020 data set, our method attains an AUROC score of $0.899$, surpassing the AST~\cite{Rudolph-WACV-2023} method by $0.004$ and the RD4AD~\cite{Deng-CVPR-2022} framework by $0.013$. On the LUNA16 benchmark, our method reaches an AUROC score of $0.634$ while RD4AD~\cite{Deng-CVPR-2022} reaches only $0.521$. The improved performance proves the benefits of introducing supervised signal in the training process.


\vspace{-0.5\baselineskip}
\subsection{Ablation Study} 
\label{sec_exp_ablation}
\vspace{-0.5\baselineskip}

\begin{figure*}[t]
\begin{center}
\includegraphics[width=0.99\linewidth]{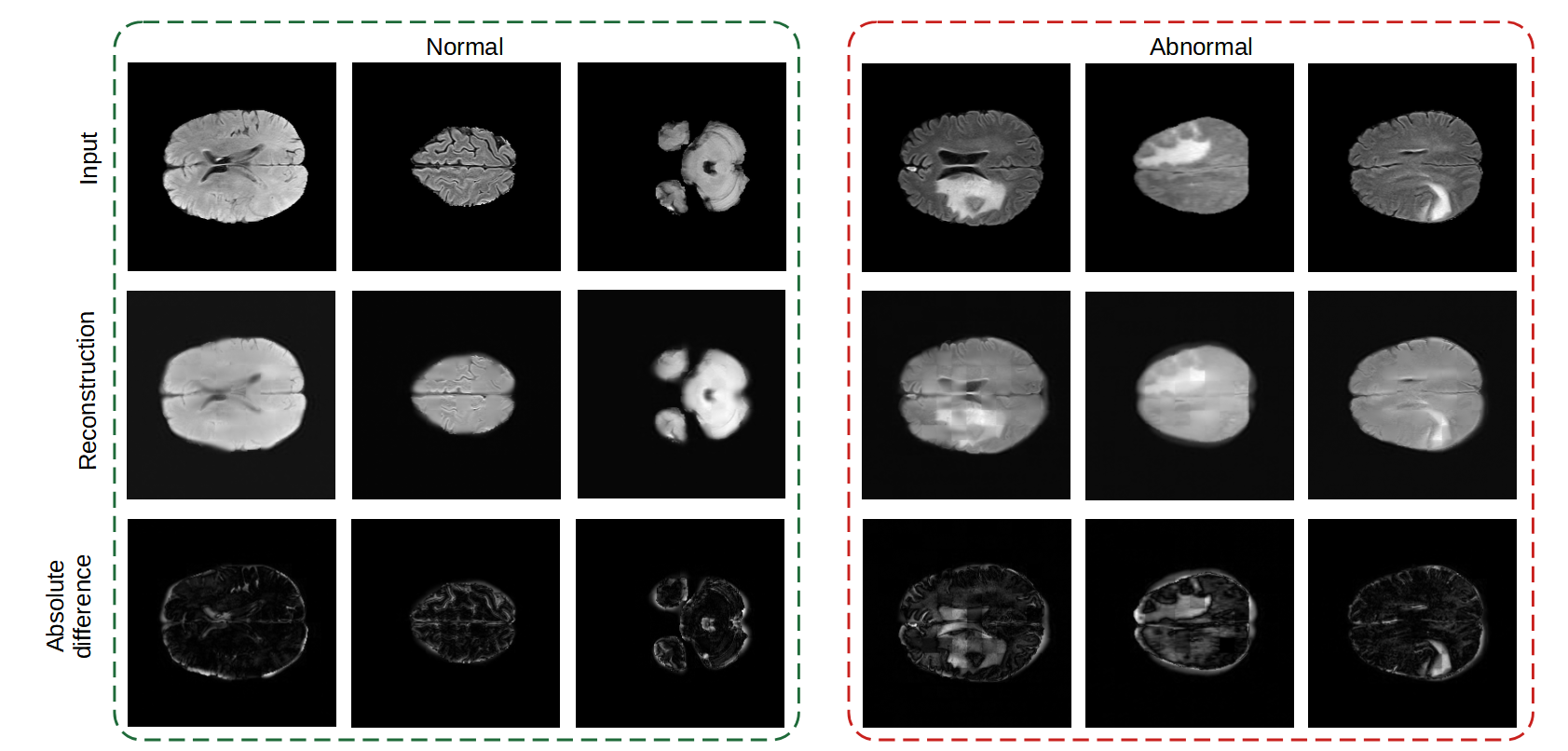}
\end{center}
\caption{Examples of reconstruction images along with the input image and the absolute difference between them obtained by applying the MAE framework on the BRATS2020~\cite{Menze-TMI-2015} data set.}\label{fig_reconstruction}
\end{figure*}

\noindent
\textbf{Anomaly classifier.} In order to assess the performance brought by the anomaly classifier, we remove it and compute the AUROC score by employing only the remaining components. When we remove the anomaly classifier, the resulting framework contains only the MAE component. In order to compute the anomaly score, we use the mean absolute error, mean squared error or structural similarity index measure (SSIM) between the reconstructed image and the original image. We compute the performance obtained by the MAE component on the BRATS2020 data set, presenting the results in Table~\ref{ablation_method}. We observe that the highest performance of $0.877$ in terms of AUROC score is obtained when SSIM is employed as the anomaly measure, showing that MAE alone is $0.012$ below our proposed framework, highlighting the significance of the anomaly classifier. In Figure~\ref{fig_reconstruction}, we show the MAE reconstruction of normal and abnormal samples from the BRATS2020 benchmark. We can easily notice that the abnormal samples have a higher reconstruction error than the normal samples.

We also ablate the choice of using the absolute difference between the reconstructed and the original image as the input for the anomaly classifier. We present the results obtained on the  BRATS2020 data set in Table~\ref{ablation_input}. We test the absolute difference ($L_1$) and the Euclidean distance ($L_2$) between the reconstructed and the original image. We also perform an experiment directly using the reconstructed image as input to the anomaly classifier. We observe that the highest performance is obtained when the absolute difference between the samples (original and reconstructed) is employed, while the lowest performance of $0.821$ is obtained when the reconstructed image is used as input to the anomaly classifier.

\noindent
\textbf{Pseudo-abnormal Module.} 
In the pseudo-abnormal module, we adjust the reconstructed samples to simulate abnormalities. We change the pixel intensities inside each selected region by multiplying the pixels with $\beta$, a random \textit{single} number drawn from a uniform distribution. Our first ablation experiment is to change the intensity of each pixel in the selected region independently (i.e. to generate a matrix of $\beta$, one for each pixel inside the bounding box). By independently changing the pixel intensity the AUROC score drops to $0.835$ from $0.899$ on the BRATS2020 data set. We conclude that independently altering each pixel inside a region is too severe, resulting in unnatural reconstruction artifacts that are not aligned with the reconstruction errors that occur when a real anomaly is encountered, therefore, the performance decreases significantly.

We also ablate the hyper-parameter $k$ (the number of pseudo-abnormal bounding boxes) which is uniformly sampled between $[\ k_1, k_2 ]\ $. We report the results obtained on the BRATS2020 data set in Table~\ref{ablation_regions}. We observe that if we use too many abnormal regions (at least $5$ per sample) the performance drops to $0.889$. We also observe that using maximum $5$ abnormal regions is slightly better than using maximum $10$ regions per sample. We carried on the experiments with the latter setting since we tuned the hyper-parameters on the validation set, where we obtained slightly better performance when using maximum $10$ regions per sample.

\begin{figure*}[t]
\begin{center}
\includegraphics[width=0.99\linewidth]{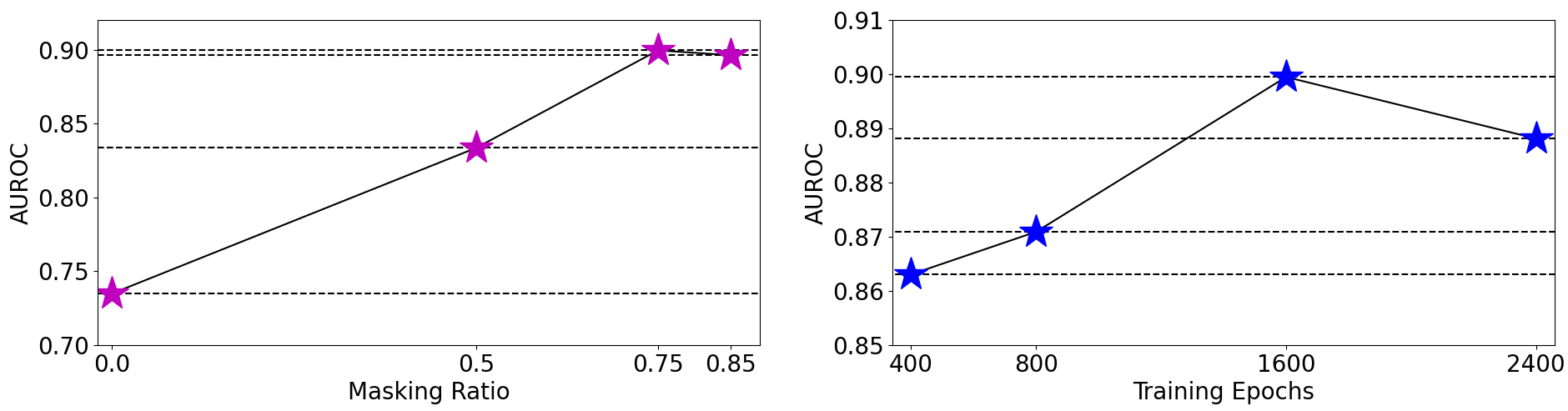}
\end{center}
\vspace{-1.5\baselineskip}
\caption{Ablation results showing the effect of the hyper-parameters (masking ratio and number of training epochs) of the MAE framework on medical image anomaly detection performance on the BRATS2020~\cite{Menze-TMI-2015} data set.}\label{fig_ablation_mae}

\end{figure*}

\noindent
\textbf{Masked Autoencoders.}
In our main experiment, we do not apply the loss on the unmasked tokens as detailed in Eq.~\eqref{eq:reconstruction_mae}. When we applied the loss on the unmasked tokens and used the reconstructed tokens instead of the original tokens, the performance drops from $0.899$ to $0.864$ in terms of AUROC on the BRATS2020 data set. Furthermore, the MAE framework has two significant hyper-parameters, namely the masking ratio $\alpha$ and the numbers of training epochs. We studied the influence of these two hyper-parameters on the medical anomaly detection performance and reported the results on the BRATS2020 data set in Figure~\ref{fig_ablation_mae}. We notice that the masking ratio equals to $0.75$ is the optimal masking ratio in order to capture representative patterns from the normal training data. We conducted an ablation experiment where we substituted the MAE framework with an Autoencoder having identical architecture (i.e. we did not mask any token and reconstruct all tokens during training) while keeping the rest of our framework unchanged. This ablated version is illustrated in the left side of Figure~\ref{fig_ablation_mae} with masking ratio equals $0$. When we replaced the MAE model with the standard AE model the performance dropped to $0.735$ in terms of AUROC on the BRATS2020 data set. To further assess the significance of the MAE framework, we removed it altogether and applied the anomaly classifier directly to the original images. This ablated version obtained an AUROC score of $0.874$ on the BRATS2020 data set. Based on the results obtained by the two ablation experiments, we conjecture that the MAE framework is a significant component of our framework.  
As illustrated in Figure~\ref{fig_ablation_mae}, we observed that training the MAE framework longer (up to $1600$ epochs) attains better performance, however after a certain point ($1600$ epochs), training MAE longer does not increase the performance of the anomaly detection framework.

\vspace{-1.2\baselineskip}
\section{Conclusions}   
\vspace{-0.9\baselineskip}
In this paper, we proposed a framework for unsupervised anomaly detection in medical images.  
We tested our framework on two data sets, namely BRATS2020~\cite{Menze-TMI-2015} and LUNA16~\cite{Setio-MIA-2017} obtaining higher performance than four state-of-the-art anomaly detection frameworks~\cite{Baur-MICCAIW-2019,Deng-CVPR-2022,Rudolph-WACV-2023,Schlegl-MIA-2019}. We performed an extensive ablation study showing the benefits of each design choice of our framework. Different from the related works~\cite{Baur-MICCAIW-2019,Deng-CVPR-2022,Rudolph-WACV-2023,Schlegl-MIA-2019}, we added supervised signal to the learning process through our pseudo-abnormal module and anomaly classifier, which helped us achieve state-of-the-art results. Our method has the advantage of not depending solely on the accuracy of the underlying model, leveraging the supervised signal obtained from the anomaly classifier. By overcoming the limitation of the previous methods (i.e. the reliance on the reconstruction model), we introduce a new challenge namely, the simulation of anomalies in healthy scans. In future work, we aim to study different approaches to simulate anomalies in medical images to further boost the performance.




 \vspace{-1.4\baselineskip}
\section*{Acknowledgements}  
 \vspace{-1.\baselineskip}
The research leading to these results has received funding from the NO Grants 2014-2021, under project ELO-Hyp contract no. 24/2020.
 \vspace{-1.3\baselineskip}



\bibliography{egbib}

\begin{thebibliography}{31}
\expandafter\ifx\csname natexlab\endcsname\relax\def\natexlab#1{#1}\fi
\providecommand{\url}[1]{\texttt{#1}}
\providecommand{\href}[2]{#2}
\providecommand{\path}[1]{#1}
\providecommand{\DOIprefix}{doi:}
\providecommand{\ArXivprefix}{arXiv:}
\providecommand{\URLprefix}{URL: }
\providecommand{\Pubmedprefix}{pmid:}
\providecommand{\doi}[1]{\href{http://dx.doi.org/#1}{\path{#1}}}
\providecommand{\Pubmed}[1]{\href{pmid:#1}{\path{#1}}}
\providecommand{\bibinfo}[2]{#2}
\ifx\xfnm\relax \def\xfnm[#1]{\unskip,\space#1}\fi
\bibitem[{Bakas et~al.(2017)Bakas, Akbari, Sotiras, Bilello, Rozycki, Kirby,
  Freymann, Farahani and Davatzikos}]{Bakas-SD-2017}
\bibinfo{author}{Bakas, S.}, \bibinfo{author}{Akbari, H.},
  \bibinfo{author}{Sotiras, A.}, \bibinfo{author}{Bilello, M.},
  \bibinfo{author}{Rozycki, M.}, \bibinfo{author}{Kirby, J.S.},
  \bibinfo{author}{Freymann, J.B.}, \bibinfo{author}{Farahani, K.},
  \bibinfo{author}{Davatzikos, C.}, \bibinfo{year}{2017}.
\newblock \bibinfo{title}{Advancing the cancer genome atlas glioma mri
  collections with expert segmentation labels and radiomic features}.
\newblock \bibinfo{journal}{Scientific data} \bibinfo{volume}{4},
  \bibinfo{pages}{1--13}.
\bibitem[{Bakas et~al.(2018)Bakas, Reyes, Jakab, Bauer, Rempfler, Crimi,
  Shinohara, Berger, Ha, Rozycki, Prastawa, Alberts, Lipkov{\'{a}}, Freymann,
  Kirby, Bilello, Fathallah{-}Shaykh, Wiest, Kirschke, Wiestler, Colen,
  Kotrotsou, LaMontagne, Marcus, Milchenko, Nazeri, Weber, Mahajan, Baid, Kwon,
  Agarwal, Alam, Albiol, Albiol, Varghese, Tuan, Arbel, Avery, B., Banerjee,
  Batchelder, Batmanghelich, Battistella, Bendszus, Benson, Bernal, Biros,
  Cabezas, Chandra, Chang and et~al.}]{Bakas-arxiv-2018}
\bibinfo{author}{Bakas, S.}, \bibinfo{author}{Reyes, M.},
  \bibinfo{author}{Jakab, A.}, \bibinfo{author}{Bauer, S.},
  \bibinfo{author}{Rempfler, M.}, \bibinfo{author}{Crimi, A.},
  \bibinfo{author}{Shinohara, R.T.}, \bibinfo{author}{Berger, C.},
  \bibinfo{author}{Ha, S.M.}, \bibinfo{author}{Rozycki, M.},
  \bibinfo{author}{Prastawa, M.}, \bibinfo{author}{Alberts, E.},
  \bibinfo{author}{Lipkov{\'{a}}, J.}, \bibinfo{author}{Freymann, J.B.},
  \bibinfo{author}{Kirby, J.S.}, \bibinfo{author}{Bilello, M.},
  \bibinfo{author}{Fathallah{-}Shaykh, H.M.}, \bibinfo{author}{Wiest, R.},
  \bibinfo{author}{Kirschke, J.}, \bibinfo{author}{Wiestler, B.},
  \bibinfo{author}{Colen, R.R.}, \bibinfo{author}{Kotrotsou, A.},
  \bibinfo{author}{LaMontagne, P.}, \bibinfo{author}{Marcus, D.S.},
  \bibinfo{author}{Milchenko, M.}, \bibinfo{author}{Nazeri, A.},
  \bibinfo{author}{Weber, M.}, \bibinfo{author}{Mahajan, A.},
  \bibinfo{author}{Baid, U.}, \bibinfo{author}{Kwon, D.},
  \bibinfo{author}{Agarwal, M.}, \bibinfo{author}{Alam, M.},
  \bibinfo{author}{Albiol, A.}, \bibinfo{author}{Albiol, A.},
  \bibinfo{author}{Varghese, A.}, \bibinfo{author}{Tuan, T.A.},
  \bibinfo{author}{Arbel, T.}, \bibinfo{author}{Avery, A.},
  \bibinfo{author}{B., P.}, \bibinfo{author}{Banerjee, S.},
  \bibinfo{author}{Batchelder, T.}, \bibinfo{author}{Batmanghelich, K.N.},
  \bibinfo{author}{Battistella, E.}, \bibinfo{author}{Bendszus, M.},
  \bibinfo{author}{Benson, E.}, \bibinfo{author}{Bernal, J.},
  \bibinfo{author}{Biros, G.}, \bibinfo{author}{Cabezas, M.},
  \bibinfo{author}{Chandra, S.}, \bibinfo{author}{Chang, Y.},
  \bibinfo{author}{et~al.}, \bibinfo{year}{2018}.
\newblock \bibinfo{title}{Identifying the best machine learning algorithms for
  brain tumor segmentation, progression assessment, and overall survival
  prediction in the {BRATS} challenge}.
\newblock \bibinfo{journal}{CoRR} \bibinfo{volume}{abs/1811.02629}.
\newblock \href{http://arxiv.org/abs/1811.02629}{{\tt arXiv:1811.02629}}.
\bibitem[{Baur et~al.(2021)Baur, Denner, Wiestler, Navab and
  Albarqouni}]{Baur-MIA-2021}
\bibinfo{author}{Baur, C.}, \bibinfo{author}{Denner, S.},
  \bibinfo{author}{Wiestler, B.}, \bibinfo{author}{Navab, N.},
  \bibinfo{author}{Albarqouni, S.}, \bibinfo{year}{2021}.
\newblock \bibinfo{title}{Autoencoders for unsupervised anomaly segmentation in
  brain mr images: A comparative study}.
\newblock \bibinfo{journal}{Medical Image Analysis} \bibinfo{volume}{69},
  \bibinfo{pages}{101952}.
\newblock \DOIprefix\doi{https://doi.org/10.1016/j.media.2020.101952}.
\bibitem[{Baur et~al.(2019)Baur, Wiestler, Albarqouni and
  Navab}]{Baur-MICCAIW-2019}
\bibinfo{author}{Baur, C.}, \bibinfo{author}{Wiestler, B.},
  \bibinfo{author}{Albarqouni, S.}, \bibinfo{author}{Navab, N.},
  \bibinfo{year}{2019}.
\newblock \bibinfo{title}{Deep autoencoding models for unsupervised anomaly
  segmentation in brain mr images}, in: \bibinfo{booktitle}{Brainlesion:
  Glioma, Multiple Sclerosis, Stroke and Traumatic Brain Injuries},
  \bibinfo{publisher}{Springer International Publishing},
  \bibinfo{address}{Cham}. pp. \bibinfo{pages}{161--169}.
\bibitem[{Baur et~al.(2020)Baur, Wiestler, Albarqouni and
  Navab}]{Baur-MICCAI-2020}
\bibinfo{author}{Baur, C.}, \bibinfo{author}{Wiestler, B.},
  \bibinfo{author}{Albarqouni, S.}, \bibinfo{author}{Navab, N.},
  \bibinfo{year}{2020}.
\newblock \bibinfo{title}{Scale-space autoencoders for unsupervised anomaly
  segmentation in brain mri}, in: \bibinfo{editor}{Martel, A.L.},
  \bibinfo{editor}{Abolmaesumi, P.}, \bibinfo{editor}{Stoyanov, D.},
  \bibinfo{editor}{Mateus, D.}, \bibinfo{editor}{Zuluaga, M.A.},
  \bibinfo{editor}{Zhou, S.K.}, \bibinfo{editor}{Racoceanu, D.},
  \bibinfo{editor}{Joskowicz, L.} (Eds.), \bibinfo{booktitle}{Medical Image
  Computing and Computer Assisted Intervention -- MICCAI 2020},
  \bibinfo{publisher}{Springer International Publishing},
  \bibinfo{address}{Cham}. pp. \bibinfo{pages}{552--561}.
\bibitem[{Bengs et~al.(2022)Bengs, Behrendt, Laves, Kr{\"u}ger, Opfer and
  Schlaefer}]{Bengs-MICAD-2022}
\bibinfo{author}{Bengs, M.}, \bibinfo{author}{Behrendt, F.},
  \bibinfo{author}{Laves, M.H.}, \bibinfo{author}{Kr{\"u}ger, J.},
  \bibinfo{author}{Opfer, R.}, \bibinfo{author}{Schlaefer, A.},
  \bibinfo{year}{2022}.
\newblock \bibinfo{title}{Unsupervised anomaly detection in 3d brain mri using
  deep learning with multi-task brain age prediction}, in:
  \bibinfo{booktitle}{Medical Imaging 2022: Computer-Aided Diagnosis},
  \bibinfo{organization}{SPIE}. pp. \bibinfo{pages}{291--295}.
\bibitem[{Chen and Konukoglu(2018)}]{Chen-MIDL-2018}
\bibinfo{author}{Chen, X.}, \bibinfo{author}{Konukoglu, E.},
  \bibinfo{year}{2018}.
\newblock \bibinfo{title}{Unsupervised detection of lesions in brain {MRI}
  using constrained adversarial auto-encoders}, in: \bibinfo{booktitle}{Medical
  Imaging with Deep Learning}.
\bibitem[{Chen et~al.(2020)Chen, You, Tezcan and Konukoglu}]{Chen-MIA-2020}
\bibinfo{author}{Chen, X.}, \bibinfo{author}{You, S.}, \bibinfo{author}{Tezcan,
  K.C.}, \bibinfo{author}{Konukoglu, E.}, \bibinfo{year}{2020}.
\newblock \bibinfo{title}{Unsupervised lesion detection via image restoration
  with a normative prior}.
\newblock \bibinfo{journal}{Medical Image Analysis} \bibinfo{volume}{64},
  \bibinfo{pages}{101713}.
\bibitem[{Deng and Li(2022)}]{Deng-CVPR-2022}
\bibinfo{author}{Deng, H.}, \bibinfo{author}{Li, X.}, \bibinfo{year}{2022}.
\newblock \bibinfo{title}{Anomaly detection via reverse distillation from
  one-class embedding}, in: \bibinfo{booktitle}{Proceedings of the IEEE/CVF
  Conference on Computer Vision and Pattern Recognition (CVPR)}, pp.
  \bibinfo{pages}{9737--9746}.
\bibitem[{Devlin et~al.(2019)Devlin, Chang, Lee and
  Toutanova}]{Devlin-ACL-2019}
\bibinfo{author}{Devlin, J.}, \bibinfo{author}{Chang, M.W.},
  \bibinfo{author}{Lee, K.}, \bibinfo{author}{Toutanova, K.},
  \bibinfo{year}{2019}.
\newblock \bibinfo{title}{{BERT}: Pre-training of deep bidirectional
  transformers for language understanding}, in: \bibinfo{booktitle}{Proceedings
  of the 2019 Conference of the North {A}merican Chapter of the Association for
  Computational Linguistics: Human Language Technologies, Volume 1 (Long and
  Short Papers)}, \bibinfo{publisher}{Association for Computational
  Linguistics}, \bibinfo{address}{Minneapolis, Minnesota}. pp.
  \bibinfo{pages}{4171--4186}.
\bibitem[{Dosovitskiy et~al.(2021)Dosovitskiy, Beyer, Kolesnikov, Weissenborn,
  Zhai, Unterthiner, Dehghani, Minderer, Heigold, Gelly, Uszkoreit and
  Houlsby}]{Dosovitskiy-ICLR-2021}
\bibinfo{author}{Dosovitskiy, A.}, \bibinfo{author}{Beyer, L.},
  \bibinfo{author}{Kolesnikov, A.}, \bibinfo{author}{Weissenborn, D.},
  \bibinfo{author}{Zhai, X.}, \bibinfo{author}{Unterthiner, T.},
  \bibinfo{author}{Dehghani, M.}, \bibinfo{author}{Minderer, M.},
  \bibinfo{author}{Heigold, G.}, \bibinfo{author}{Gelly, S.},
  \bibinfo{author}{Uszkoreit, J.}, \bibinfo{author}{Houlsby, N.},
  \bibinfo{year}{2021}.
\newblock \bibinfo{title}{An image is worth 16x16 words: Transformers for image
  recognition at scale}, in: \bibinfo{booktitle}{International Conference on
  Learning Representations}.
\bibitem[{Han et~al.(2021)Han, Rundo, Murao, Noguchi, Shimahara, Milacski,
  Koshino, Sala, Nakayama and Satoh}]{Han-BMCB-2021}
\bibinfo{author}{Han, C.}, \bibinfo{author}{Rundo, L.}, \bibinfo{author}{Murao,
  K.}, \bibinfo{author}{Noguchi, T.}, \bibinfo{author}{Shimahara, Y.},
  \bibinfo{author}{Milacski, Z.{\'A}.}, \bibinfo{author}{Koshino, S.},
  \bibinfo{author}{Sala, E.}, \bibinfo{author}{Nakayama, H.},
  \bibinfo{author}{Satoh, S.}, \bibinfo{year}{2021}.
\newblock \bibinfo{title}{Madgan: Unsupervised medical anomaly detection gan
  using multiple adjacent brain mri slice reconstruction}.
\newblock \bibinfo{journal}{BMC bioinformatics} \bibinfo{volume}{22},
  \bibinfo{pages}{1--20}.
\bibitem[{He et~al.(2022)He, Chen, Xie, Li, Doll\'ar and
  Girshick}]{He-CVPR-2022}
\bibinfo{author}{He, K.}, \bibinfo{author}{Chen, X.}, \bibinfo{author}{Xie,
  S.}, \bibinfo{author}{Li, Y.}, \bibinfo{author}{Doll\'ar, P.},
  \bibinfo{author}{Girshick, R.}, \bibinfo{year}{2022}.
\newblock \bibinfo{title}{Masked autoencoders are scalable vision learners},
  in: \bibinfo{booktitle}{Proceedings of the IEEE/CVF Conference on Computer
  Vision and Pattern Recognition (CVPR)}, pp. \bibinfo{pages}{16000--16009}.
\bibitem[{Huang et~al.(2022)Huang, Xu, Li, Baevski, Auli, Galuba, Metze and
  Feichtenhofer}]{Huang-NIPS-2022}
\bibinfo{author}{Huang, P.Y.}, \bibinfo{author}{Xu, H.}, \bibinfo{author}{Li,
  J.}, \bibinfo{author}{Baevski, A.}, \bibinfo{author}{Auli, M.},
  \bibinfo{author}{Galuba, W.}, \bibinfo{author}{Metze, F.},
  \bibinfo{author}{Feichtenhofer, C.}, \bibinfo{year}{2022}.
\newblock \bibinfo{title}{Masked autoencoders that listen}, in:
  \bibinfo{booktitle}{NeurIPS}.
\bibitem[{Jia and Chen(2020)}]{Zheshu-ACCESS-2020}
\bibinfo{author}{Jia, Z.}, \bibinfo{author}{Chen, D.}, \bibinfo{year}{2020}.
\newblock \bibinfo{title}{Brain tumor identification and classification of mri
  images using deep learning techniques}.
\newblock \bibinfo{journal}{IEEE Access} ,
  \bibinfo{pages}{1--1}\DOIprefix\doi{10.1109/ACCESS.2020.3016319}.
\bibitem[{Kascenas et~al.(2022)Kascenas, Pugeault and
  O'Neil}]{Kascenas-MIDL-2022}
\bibinfo{author}{Kascenas, A.}, \bibinfo{author}{Pugeault, N.},
  \bibinfo{author}{O'Neil, A.Q.}, \bibinfo{year}{2022}.
\newblock \bibinfo{title}{Denoising autoencoders for unsupervised anomaly
  detection in brain {MRI}}, in: \bibinfo{booktitle}{Medical Imaging with Deep
  Learning}.
\bibitem[{Loshchilov and Hutter(2019)}]{Loshchilov-ICLR-2019}
\bibinfo{author}{Loshchilov, I.}, \bibinfo{author}{Hutter, F.},
  \bibinfo{year}{2019}.
\newblock \bibinfo{title}{Decoupled weight decay regularization}, in:
  \bibinfo{booktitle}{International Conference on Learning Representations}.
\bibitem[{Menze et~al.(2015)Menze, Jakab, Bauer, Kalpathy-Cramer, Farahani,
  Kirby, Burren, Porz, Slotboom, Wiest, Lanczi, Gerstner, Weber, Arbel, Avants,
  Ayache, Buendia, Collins, Cordier, Corso, Criminisi, Das, Delingette,
  Demiralp, Durst, Dojat, Doyle, Festa, Forbes, Geremia, Glocker, Golland, Guo,
  Hamamci, Iftekharuddin, Jena, John, Konukoglu, Lashkari, Mariz, Meier,
  Pereira, Precup, Price, Raviv, Reza, Ryan, Sarikaya, Schwartz, Shin, Shotton,
  Silva, Sousa, Subbanna, Szekely, Taylor, Thomas, Tustison, Unal, Vasseur,
  Wintermark, Ye, Zhao, Zhao, Zikic, Prastawa, Reyes and
  Van~Leemput}]{Menze-TMI-2015}
\bibinfo{author}{Menze, B.H.}, \bibinfo{author}{Jakab, A.},
  \bibinfo{author}{Bauer, S.}, \bibinfo{author}{Kalpathy-Cramer, J.},
  \bibinfo{author}{Farahani, K.}, \bibinfo{author}{Kirby, J.},
  \bibinfo{author}{Burren, Y.}, \bibinfo{author}{Porz, N.},
  \bibinfo{author}{Slotboom, J.}, \bibinfo{author}{Wiest, R.},
  \bibinfo{author}{Lanczi, L.}, \bibinfo{author}{Gerstner, E.},
  \bibinfo{author}{Weber, M.A.}, \bibinfo{author}{Arbel, T.},
  \bibinfo{author}{Avants, B.B.}, \bibinfo{author}{Ayache, N.},
  \bibinfo{author}{Buendia, P.}, \bibinfo{author}{Collins, D.L.},
  \bibinfo{author}{Cordier, N.}, \bibinfo{author}{Corso, J.J.},
  \bibinfo{author}{Criminisi, A.}, \bibinfo{author}{Das, T.},
  \bibinfo{author}{Delingette, H.}, \bibinfo{author}{Demiralp, Ã.},
  \bibinfo{author}{Durst, C.R.}, \bibinfo{author}{Dojat, M.},
  \bibinfo{author}{Doyle, S.}, \bibinfo{author}{Festa, J.},
  \bibinfo{author}{Forbes, F.}, \bibinfo{author}{Geremia, E.},
  \bibinfo{author}{Glocker, B.}, \bibinfo{author}{Golland, P.},
  \bibinfo{author}{Guo, X.}, \bibinfo{author}{Hamamci, A.},
  \bibinfo{author}{Iftekharuddin, K.M.}, \bibinfo{author}{Jena, R.},
  \bibinfo{author}{John, N.M.}, \bibinfo{author}{Konukoglu, E.},
  \bibinfo{author}{Lashkari, D.}, \bibinfo{author}{Mariz, J.A.},
  \bibinfo{author}{Meier, R.}, \bibinfo{author}{Pereira, S.},
  \bibinfo{author}{Precup, D.}, \bibinfo{author}{Price, S.J.},
  \bibinfo{author}{Raviv, T.R.}, \bibinfo{author}{Reza, S.M.S.},
  \bibinfo{author}{Ryan, M.}, \bibinfo{author}{Sarikaya, D.},
  \bibinfo{author}{Schwartz, L.}, \bibinfo{author}{Shin, H.C.},
  \bibinfo{author}{Shotton, J.}, \bibinfo{author}{Silva, C.A.},
  \bibinfo{author}{Sousa, N.}, \bibinfo{author}{Subbanna, N.K.},
  \bibinfo{author}{Szekely, G.}, \bibinfo{author}{Taylor, T.J.},
  \bibinfo{author}{Thomas, O.M.}, \bibinfo{author}{Tustison, N.J.},
  \bibinfo{author}{Unal, G.}, \bibinfo{author}{Vasseur, F.},
  \bibinfo{author}{Wintermark, M.}, \bibinfo{author}{Ye, D.H.},
  \bibinfo{author}{Zhao, L.}, \bibinfo{author}{Zhao, B.},
  \bibinfo{author}{Zikic, D.}, \bibinfo{author}{Prastawa, M.},
  \bibinfo{author}{Reyes, M.}, \bibinfo{author}{Van~Leemput, K.},
  \bibinfo{year}{2015}.
\newblock \bibinfo{title}{The multimodal brain tumor image segmentation
  benchmark (brats)}.
\newblock \bibinfo{journal}{IEEE Transactions on Medical Imaging}
  \bibinfo{volume}{34}, \bibinfo{pages}{1993--2024}.
\newblock \DOIprefix\doi{10.1109/TMI.2014.2377694}.
\bibitem[{Mohsen et~al.(2018)Mohsen, El-Dahshan, El-Horbaty and
  Salem}]{Mohsen-FCIJ-2018}
\bibinfo{author}{Mohsen, H.}, \bibinfo{author}{El-Dahshan, E.S.A.},
  \bibinfo{author}{El-Horbaty, E.S.M.}, \bibinfo{author}{Salem, A.B.M.},
  \bibinfo{year}{2018}.
\newblock \bibinfo{title}{Classification using deep learning neural networks
  for brain tumors}.
\newblock \bibinfo{journal}{Future Computing and Informatics Journal}
  \bibinfo{volume}{3}, \bibinfo{pages}{68--71}.
\bibitem[{Pinaya et~al.(2022a)Pinaya, Tudosiu, Gray, Rees, Nachev, Ourselin and
  Cardoso}]{Pinaya-MIA-2022}
\bibinfo{author}{Pinaya, W.H.}, \bibinfo{author}{Tudosiu, P.D.},
  \bibinfo{author}{Gray, R.}, \bibinfo{author}{Rees, G.},
  \bibinfo{author}{Nachev, P.}, \bibinfo{author}{Ourselin, S.},
  \bibinfo{author}{Cardoso, M.J.}, \bibinfo{year}{2022}a.
\newblock \bibinfo{title}{Unsupervised brain imaging 3d anomaly detection and
  segmentation with transformers}.
\newblock \bibinfo{journal}{Medical Image Analysis} \bibinfo{volume}{79},
  \bibinfo{pages}{102475}.
\bibitem[{Pinaya et~al.(2022b)Pinaya, Graham, Gray, da~Costa, Tudosiu, Wright,
  Mah, MacKinnon, Teo, Jager, Werring, Rees, Nachev, Ourselin and
  Cardoso}]{Pinaya-MICCAI-2022}
\bibinfo{author}{Pinaya, W.H.L.}, \bibinfo{author}{Graham, M.S.},
  \bibinfo{author}{Gray, R.}, \bibinfo{author}{da~Costa, P.F.},
  \bibinfo{author}{Tudosiu, P.D.}, \bibinfo{author}{Wright, P.},
  \bibinfo{author}{Mah, Y.H.}, \bibinfo{author}{MacKinnon, A.D.},
  \bibinfo{author}{Teo, J.T.}, \bibinfo{author}{Jager, R.},
  \bibinfo{author}{Werring, D.}, \bibinfo{author}{Rees, G.},
  \bibinfo{author}{Nachev, P.}, \bibinfo{author}{Ourselin, S.},
  \bibinfo{author}{Cardoso, M.J.}, \bibinfo{year}{2022}b.
\newblock \bibinfo{title}{Fast unsupervised brain anomaly detection
  and segmentation with diffusion models}, in: \bibinfo{editor}{Wang, L.},
  \bibinfo{editor}{Dou, Q.}, \bibinfo{editor}{Fletcher, P.T.},
  \bibinfo{editor}{Speidel, S.}, \bibinfo{editor}{Li, S.} (Eds.),
  \bibinfo{booktitle}{Medical Image Computing and Computer Assisted
  Intervention -- MICCAI 2022}, \bibinfo{publisher}{Springer Nature
  Switzerland}, \bibinfo{address}{Cham}. pp. \bibinfo{pages}{705--714}.
\bibitem[{Rudolph et~al.(2023)Rudolph, Wehrbein, Rosenhahn and
  Wandt}]{Rudolph-WACV-2023}
\bibinfo{author}{Rudolph, M.}, \bibinfo{author}{Wehrbein, T.},
  \bibinfo{author}{Rosenhahn, B.}, \bibinfo{author}{Wandt, B.},
  \bibinfo{year}{2023}.
\newblock \bibinfo{title}{Asymmetric student-teacher networks for industrial
  anomaly detection}, in: \bibinfo{booktitle}{Winter Conference on Applications
  of Computer Vision (WACV)}.
\bibitem[{Schlegl et~al.(2017)Schlegl, Seeb{\"o}ck, Waldstein, Schmidt-Erfurth
  and Langs}]{Schlegl-IPMI-2017}
\bibinfo{author}{Schlegl, T.}, \bibinfo{author}{Seeb{\"o}ck, P.},
  \bibinfo{author}{Waldstein, S.M.}, \bibinfo{author}{Schmidt-Erfurth, U.},
  \bibinfo{author}{Langs, G.}, \bibinfo{year}{2017}.
\newblock \bibinfo{title}{Unsupervised anomaly detection with generative
  adversarial networks to guide marker discovery}, in:
  \bibinfo{editor}{Niethammer, M.}, \bibinfo{editor}{Styner, M.},
  \bibinfo{editor}{Aylward, S.}, \bibinfo{editor}{Zhu, H.},
  \bibinfo{editor}{Oguz, I.}, \bibinfo{editor}{Yap, P.T.},
  \bibinfo{editor}{Shen, D.} (Eds.), \bibinfo{booktitle}{Information Processing
  in Medical Imaging}, \bibinfo{publisher}{Springer International Publishing},
  \bibinfo{address}{Cham}. pp. \bibinfo{pages}{146--157}.
\bibitem[{Schlegl et~al.(2019)Schlegl, Seeböck, Waldstein, Langs and
  Schmidt-Erfurth}]{Schlegl-MIA-2019}
\bibinfo{author}{Schlegl, T.}, \bibinfo{author}{Seeböck, P.},
  \bibinfo{author}{Waldstein, S.M.}, \bibinfo{author}{Langs, G.},
  \bibinfo{author}{Schmidt-Erfurth, U.}, \bibinfo{year}{2019}.
\newblock \bibinfo{title}{f-anogan: Fast unsupervised anomaly detection with
  generative adversarial networks}.
\newblock \bibinfo{journal}{Medical Image Analysis} \bibinfo{volume}{54},
  \bibinfo{pages}{30--44}.
\bibitem[{Setio et~al.(2017)Setio, Traverso, {de Bel}, Berens, van~den Bogaard,
  Cerello, Chen, Dou, Fantacci, Geurts, van~der Gugten, Heng, Jansen, {de
  Kaste}, Kotov, Lin, Manders, Sóñora-Mengana, García-Naranjo,
  Papavasileiou, Prokop, Saletta, Schaefer-Prokop, Scholten, Scholten, Snoeren,
  Torres, Vandemeulebroucke, Walasek, Zuidhof, van Ginneken and
  Jacobs}]{Setio-MIA-2017}
\bibinfo{author}{Setio, A.A.A.}, \bibinfo{author}{Traverso, A.},
  \bibinfo{author}{{de Bel}, T.}, \bibinfo{author}{Berens, M.S.},
  \bibinfo{author}{van~den Bogaard, C.}, \bibinfo{author}{Cerello, P.},
  \bibinfo{author}{Chen, H.}, \bibinfo{author}{Dou, Q.},
  \bibinfo{author}{Fantacci, M.E.}, \bibinfo{author}{Geurts, B.},
  \bibinfo{author}{van~der Gugten, R.}, \bibinfo{author}{Heng, P.A.},
  \bibinfo{author}{Jansen, B.}, \bibinfo{author}{{de Kaste}, M.M.},
  \bibinfo{author}{Kotov, V.}, \bibinfo{author}{Lin, J.Y.H.},
  \bibinfo{author}{Manders, J.T.}, \bibinfo{author}{Sóñora-Mengana, A.},
  \bibinfo{author}{García-Naranjo, J.C.}, \bibinfo{author}{Papavasileiou, E.},
  \bibinfo{author}{Prokop, M.}, \bibinfo{author}{Saletta, M.},
  \bibinfo{author}{Schaefer-Prokop, C.M.}, \bibinfo{author}{Scholten, E.T.},
  \bibinfo{author}{Scholten, L.}, \bibinfo{author}{Snoeren, M.M.},
  \bibinfo{author}{Torres, E.L.}, \bibinfo{author}{Vandemeulebroucke, J.},
  \bibinfo{author}{Walasek, N.}, \bibinfo{author}{Zuidhof, G.C.},
  \bibinfo{author}{van Ginneken, B.}, \bibinfo{author}{Jacobs, C.},
  \bibinfo{year}{2017}.
\newblock \bibinfo{title}{Validation, comparison, and combination of algorithms
  for automatic detection of pulmonary nodules in computed tomography images:
  The luna16 challenge}.
\newblock \bibinfo{journal}{Medical Image Analysis} \bibinfo{volume}{42},
  \bibinfo{pages}{1--13}.
\bibitem[{Shamshad et~al.(2022)Shamshad, Khan, Zamir, Khan, Hayat, Khan and
  Fu}]{Shamshad-arxiv-2022}
\bibinfo{author}{Shamshad, F.}, \bibinfo{author}{Khan, S.},
  \bibinfo{author}{Zamir, S.W.}, \bibinfo{author}{Khan, M.H.},
  \bibinfo{author}{Hayat, M.}, \bibinfo{author}{Khan, F.S.},
  \bibinfo{author}{Fu, H.}, \bibinfo{year}{2022}.
\newblock \bibinfo{title}{Transformers in medical imaging: A survey}.
\bibitem[{Shvetsova et~al.(2021)Shvetsova, Bakker, Fedulova, Schulz and
  Dylov}]{Shvetsova-ACCESS-2021}
\bibinfo{author}{Shvetsova, N.}, \bibinfo{author}{Bakker, B.},
  \bibinfo{author}{Fedulova, I.}, \bibinfo{author}{Schulz, H.},
  \bibinfo{author}{Dylov, D.V.}, \bibinfo{year}{2021}.
\newblock \bibinfo{title}{Anomaly detection in medical imaging with deep
  perceptual autoencoders}.
\newblock \bibinfo{journal}{IEEE Access} \bibinfo{volume}{9},
  \bibinfo{pages}{118571--118583}.
\newblock \DOIprefix\doi{10.1109/ACCESS.2021.3107163}.
\bibitem[{Song et~al.(2020)Song, Zou, Zhou, Huang, Shao, Yuan, Gou, Jin, Wang,
  Chen et~al.}]{Song-NC-2020}
\bibinfo{author}{Song, Z.}, \bibinfo{author}{Zou, S.}, \bibinfo{author}{Zhou,
  W.}, \bibinfo{author}{Huang, Y.}, \bibinfo{author}{Shao, L.},
  \bibinfo{author}{Yuan, J.}, \bibinfo{author}{Gou, X.}, \bibinfo{author}{Jin,
  W.}, \bibinfo{author}{Wang, Z.}, \bibinfo{author}{Chen, X.}, et~al.,
  \bibinfo{year}{2020}.
\newblock \bibinfo{title}{Clinically applicable histopathological diagnosis
  system for gastric cancer detection using deep learning}.
\newblock \bibinfo{journal}{Nature communications} \bibinfo{volume}{11},
  \bibinfo{pages}{4294}.
\bibitem[{Tong et~al.(2022)Tong, Song, Wang and Wang}]{Tong-NIPS-2022}
\bibinfo{author}{Tong, Z.}, \bibinfo{author}{Song, Y.}, \bibinfo{author}{Wang,
  J.}, \bibinfo{author}{Wang, L.}, \bibinfo{year}{2022}.
\newblock \bibinfo{title}{Video{MAE}: Masked autoencoders are data-efficient
  learners for self-supervised video pre-training}, in:
  \bibinfo{booktitle}{Advances in Neural Information Processing Systems}.
\bibitem[{Wolleb et~al.(2022)Wolleb, Bieder, Sandk{\"u}hler and
  Cattin}]{Wolleb-MICCAI-2022}
\bibinfo{author}{Wolleb, J.}, \bibinfo{author}{Bieder, F.},
  \bibinfo{author}{Sandk{\"u}hler, R.}, \bibinfo{author}{Cattin, P.C.},
  \bibinfo{year}{2022}.
\newblock \bibinfo{title}{Diffusion models for medical anomaly detection}, in:
  \bibinfo{editor}{Wang, L.}, \bibinfo{editor}{Dou, Q.},
  \bibinfo{editor}{Fletcher, P.T.}, \bibinfo{editor}{Speidel, S.},
  \bibinfo{editor}{Li, S.} (Eds.), \bibinfo{booktitle}{Medical Image Computing
  and Computer Assisted Intervention -- MICCAI 2022},
  \bibinfo{publisher}{Springer Nature Switzerland}, \bibinfo{address}{Cham}.
  pp. \bibinfo{pages}{35--45}.
\bibitem[{Zimmerer et~al.(2019)Zimmerer, Kohl, Petersen, Isensee and
  Maier-Hein}]{Zimmerer-MIDL-2019}
\bibinfo{author}{Zimmerer, D.}, \bibinfo{author}{Kohl, S.},
  \bibinfo{author}{Petersen, J.}, \bibinfo{author}{Isensee, F.},
  \bibinfo{author}{Maier-Hein, K.}, \bibinfo{year}{2019}.
\newblock \bibinfo{title}{Context-encoding variational autoencoder for
  unsupervised anomaly detection}.

\end{thebibliography}
\bibliographystyle{elsarticle-harv}

\end{document}